\documentclass{PoS}

\def \etal   {\hbox{\it et~al.\/}}

\title{Polarization and SEDs from Microlensing of Circumstellar Envelopes}

\ShortTitle{Microlensing of Circumstellar Envelopes}

\author{\speaker{Richard Ignace}\\
        Department of Physics, Astronomy, \& Geology,
	East Tennessee State University\\
        E-mail: \email{ignace@etsu.edu}}

\author{Jon E.\ Bjorkman\\
        Department of Physics and Astronomy, University of Toledo\\
        E-mail: \email{jon@physics.utoledo.edu}}

\author{Christina Bunker\thanks{Southeastern Association for Research in Astronomy (SARA) NSF-REU Summer Intern}\\
        Department of Physics and Astronomy, SUNY Stony Brook\\
        E-mail: \email{christina.bunker@gmail.com}}

\abstract{
Microlensing surveys have proven to be tremendously fruitful in providing
valuable data products for many fields of astrophysics, from eclipse
lightcurves for substellar candidates to limb darkening in stellar
atmospheres.  We report on a program of modeling observables from
microlensing of circumstellar envelopes, particularly those of red
giant stars that are the most likely to show finite source effects.
We will summarize work for how polarization light curves can be used to
infer envelope properties and will describe recent modeling of the
time dependent spectral energy distributions (SEDs) for microlensing of
dusty winds.  One of the most exciting developments is the possibility
of measuring variable polarization from microlensing in a suitable
source using the RINGO polarimeter at La Palma.  Also quite interesting
is the possibility of probing a dusty wind using IRAC data for a suitable
source in the event that Spitzer has a ``warm'' cycle.
          }

\FullConference{The Manchester Microlensing Conference: The 12th International Conference and ANGLES Microlensing Workshop\\
		 January 21-25 2008\\
		 Manchester, UK}

\begin{document}

\section{Introduction}

Galactic microlensing has produced a number of results for stellar
atmospheres as an important offshoot of the original intent to understand
dark matter in the Galactic halo.  Among these have been applications
to stellar atmospheres (e.g., Gould 2001) such as limb darkening
(e.g., Albrow \etal\ 1999, 2001; Fields \etal\ 2003; Abe \etal\ 2003;
Cassan \etal\ 2006) and chromospheres in red giant stars (Afonso \etal\
2001; Cassan \etal\ 2004; Thurl \etal\ 2006).  The possibility for what
amounts to mapping stellar atmospheres using microlensing arises when
the lens and background source have a transit or near transit event.
The more spectacular events, and consequently the ones most suitable
for recovering the intensity profile of the source atmosphere, occur
for binary lenses when the source has a caustic crossing.

Researchers have been pushing the envelope of possibilities for studying
stellar atmospheres through considerations of more complex effects
that might be detectable.  These include stellar rotation (Gould 1997),
stellar spots (Heyrovsky \& Sasselov 2000; Han \etal\ 2000; Hendry \etal\
2002), stellar pulsation (Bryce 2001), and differential stellar rotation
(Hendry \& Ignace, in prep).  In contrast, our group has focused largely
on the prospects of constraining the structure of {\em circumstellar}
envelopes through microlensing events.

There are several motivations for considering the influence of
circumstellar emissions in the spectra and photometry from source transit
events.  First, the most likely targets to show finite source effects are
red giant stars, and red giants have significant winds.  Their mass-loss
rates $\dot{M}$ have values of around $10^{-10}$ to $10^{-8} M_\odot/$yr,
all the way to $10^{-5} M_\odot/$yr for rare asymptotic giant branch (AGB)
stars (Dupree 1986, Willson 2000).  Second, the opportunity to infer the
intensity profile of these sources through microlensing translates to
mapping the wind flow structure.  The red giants are interesting because
the driving of their winds is still not well-understood.  Certainly the
AGB winds are dust-driven (Netzer \& Elitzur 1993).  The lower $\dot{M}$
winds of red giants may be driven by some combination of molecular
opacity and dust (Jorgensen \& Johnson 1992).  Finally, an extended
envelope simply means a larger source thus making a transit more likely,
which may lead to additional constraints on the nature of the lens.

What kinds of questions about the atmospheres and winds of red giants
could be addressed through microlensing?  Although dust is known to
play a role (even a dominant role in the case of AGBs) in the wind
driving, dust condensation temperatures are typically below 2000~K,
whereas the red giant atmospheres are around 3000~K or more.  This means
dust formation occurs in the wind; however, the dust formation radius
is not well-constrained observationally.  Additionally, there is a
well-known division in the Hertzsprung-Russell Diagram (HRD) known as the
``Linsky-Haisch dividing line'' among giant stars between those that
show X-ray emissions and those that do not (Linsky \& Haisch 1979);
the cause for this division is still being considered (Suzuki 2007).
Both the ionized chromospheres of red
giants and the stellar winds with molecular or dust opacity offer {\em
scattering} opacity sources that lead to polarization, hence our group
has been exploring variable polarization signals that can arise from
microlensing transit events.
This contribution reviews the main results from the polarization
calculations and prospects for detections and presents new results for
the influence of microlensing on the spectral energy distributions (SEDs)
of dusty wind sources.  Our team has also studied emission line
shape variations for lines that form in circumstellar envelopes
(Ignace \& Hendry 1999; Bryce \etal\ 2003; Hendry \etal\ 2006).  Such
variations may be realized in appropriate lines for red giant winds
or in supernovae; however, space does not permit a discussion of these
models here.

\begin{figure}
\includegraphics[width=.6\textwidth]{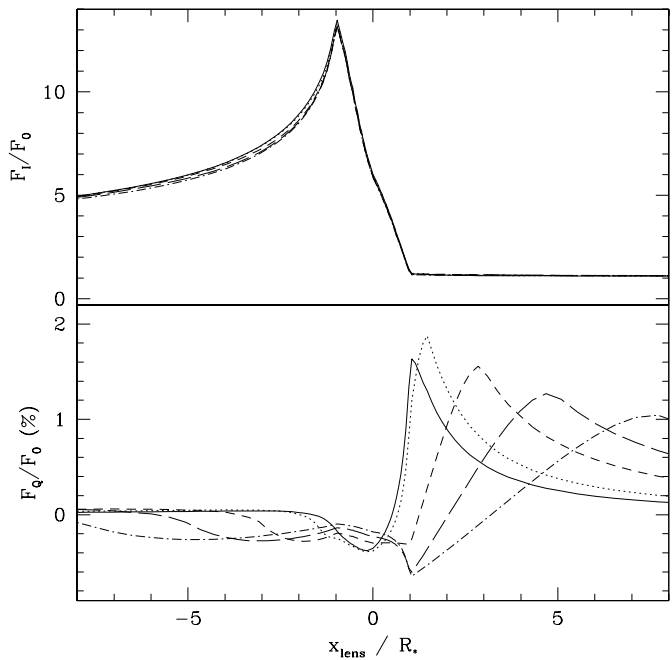}
\caption{Stokes-I and Q fluxes normalized to the unlensed value $F_0$ and
plotted for different positions relative to a straight line fold caustic
(from Ignace \etal\ 2006).
Negative values of $x_{\rm lens}$ are interior to the caustic; positive
are exterior.  The Thomson scattering optical depth is $\tau_{\rm sc}
= 0.1$.  Each curve is for a different hole radius $R_{\rm h} = 1.0,
1.5, 3.0, 5.0,$ and $8.0 R_\ast$, with larger cavities giving variations
spread over a larger $x_{\rm lens}$ interval.
}
\label{fig1}
\end{figure}

\section{Variable Polarization from Microlensing}

For an unresolved but spherically
symmetric source with scattering opacity, the total polarization will
sum to zero.  Although there is a non-zero polarized intensity
along any given ray through the source medium, the polarizations
from all the rays combine vectorially (e.g., using the Stokes notation $I,
Q, U, V$) such that the integration of contributions from all rays must
sum to zero.  A net polarization can only result from some asymmetry
in the system, possibly a non-spherical envelope (Brown \& McLean 1977) or
asymmetric illumination of scattering opacity (Al-Malki \etal\ 1999).
Yet even in spherical symmetry, microlensing leads to a net polarization
because it introduces an asymmetry into the system by virtue of selective
magnification.

Schneider \& Wagoner (1987) initiated ideas about variable polarization
during microlensing events in their study of pseudo-photospheres
of SNe.  As galactic microlensing began to produce detections, and
especially finite source events, Simmons \etal\ (1995ab) and Agol (1996)
(and more recently Yoshida 2006) considered polarimetric signals from
stellar atmospheres, while Simmons \etal\ (2002) and Ignace \etal\
(2006) have extended applications to circumstellar envelopes.

A key motivation for considering circumstellar envelopes is the
possibility of determining the dust condensation radius.  To explore
such effects, Simmons \etal\ (2002) and Ignace \etal\ (2006) modeled
the envelope as having an inner radius ($R_{\rm h}$) for a ``hole''
of scattering opacity.  Certainly gas exists between the photosphere
at $R_\ast$ and $R_{\rm h}$, only that zone was assumed not to contribute
to scattered light.  Those models further assumed optically thin dipole
scattering for simplicity.  Example results for caustic crossing events
are shown in Figure~\ref{fig1}.  The upper panel shows the normalized
Stokes-$I$ flux as the source moves from inside a straight line fold
caustic to the outside.  The lower panel shows the polarized emission.
Note that the geometry arrangement ensures that Stokes-$U$ is zero.
The change in sign of $Q$ indicates a position angle flip between
parallel ($+$) and perpendicular ($-$) orientations to the caustic line.
Different curves are for different ``hole radii'' as labeled.  It is clear
that for binary lensing, quite large polarizations could result; however,
this will depend on the envelope optical depth.  In addition, the models
in figure~\ref{fig1} ignore the potential for any polarization contribution from
the photosphere or chromosphere, or molecular Rayleigh scattering
that might exist in the ``hole'' region.  All of these effects need to
be explored more carefully using realistic stellar atmospheres.

It is worth noting that a general feature of polarization from
microlensing is a measurement of the polarization position angle in
the sky.  This position angle provides information about the {\em vector}
relative proper motion between the source and lens, information that in
principle could prove useful in helping to constrain the lens location
and its population.

It should also be pointed out that there is a claim for the detection of
variable polarization from microlensing for a quasar (Chae \etal\ 2001).
At the time of this writing, there has been no attempt to measure the
effect in galactic microlensing; however, Iain Steele of the University of
Liverpool-John Moores has agreed to provide discretionary time using the
polarimeter RINGO (Steele \etal\ 2006) to measure broad band polarization
in a suitable microlensing event.

\begin{figure}
\includegraphics[width=.6\textwidth]{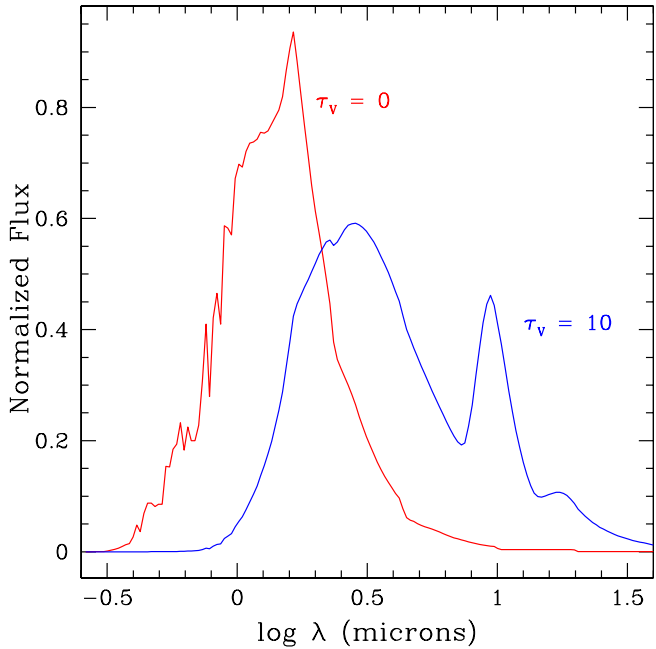}
\caption{Left is the input Kurucz model spectrum appropriate for a red
giant star; right shows the strongly redistributed emergent spectrum
when this star has a thick dusty wind of $\tau_V = 10$.}
\label{fig2}
\end{figure}

\section{Variable SEDs from Dusty Envelopes}

Although dust may be present in the lower mass loss winds of typical
red giants, it plays a dominant role in the wind driving for the high
mass-loss rate AGB stars with $\dot{M} \approx 10^{-7} M_\odot/$yr
or more.  The presence of dust can alter or dominate the SEDs
of red giants and AGB stars.  A Monte Carlo radiative transfer algorithm
has been used to model the SEDs of dusty winds during
microlensing.  The models assume spherical
symmetry and a density profile $\rho \propto r^{-2}$.  Radiative
equilibrium is enforced following Bjorkman \& Wood (2001).  As a result,
a self-consistent temperature profile can be found for the dust.  For the
wind, the free parameters include the envelope optical depth $\tau_V$
(at 5500~\AA) and the dust condensation radius $R_{\rm h}$.  In principle,
different grain species could also be considered; however, a standard
interstellar dust opacity was adopted (Kim \etal\ 1994).  The remaining
free parameter is then the Einstein radius in the source plane $\hat{r}_E$
relative to $R_{\rm h}$.

Figure~\ref{fig2} illustrates the effect that dust can have on the
stellar SED.  There is no lensing in this example.  The underlying stellar
atmosphere is a Kurucz model at $T_{\rm eff} = 3500$~K (1992) shown for
the case of $\tau_V=0$.  Also shown is a substantially dust-enshrouded
star at $\tau_V=10$, displaying a radically different SED.  The prominent
IR bump near 10~$\mu$m is dust silicate emission.

Key to understanding chromatic effects from microlensing transits is that
the ``effective'' source size is a function of wavelength, $R_\lambda$.
It is this effect that has allowed for the derivation of limb darkening
from photometry and the probing of red giant chromospheres in H$\alpha$.
For a wind the density is generally power-law like, and the opacity is
roughly power-law except in certain wavelength intervals.  The combination
can lead to a quite substantial range in $R_\lambda$ values as a function
of wavelength.

\begin{figure}
\includegraphics[width=.6\textwidth]{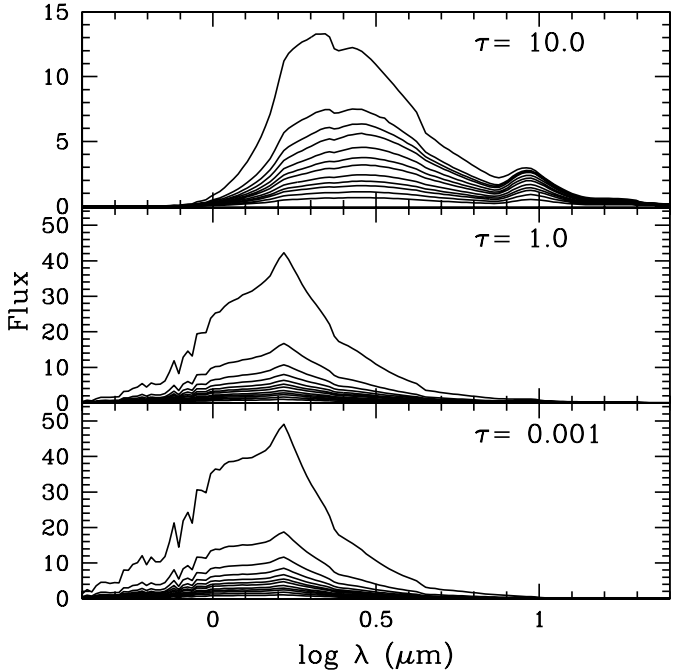}
\caption{Simulated spectral variations for dusty envelopes with three
different dust optical depths as indicated.  Each has $R_{\rm h}=8 R_\ast$
and $\hat{r}_E=10 R_{\rm h}$.  The unlensed spectrum is normalized to
have unit wavelength integrated flux.  Different curves are for different
lens positions, with the most magnified cases corresponding to $x_{\rm lens}
= 0.1R_{\rm h}$.
} 
\label{fig3}
\end{figure}

Figure~\ref{fig3} displays one suite of model results for microlensing
of dusty envelopes.  This log-linear plot shows spectra in wavelength.
The panels are for envelopes of different dust optical depths as labeled
(again evaluated at 5500~\AA).  The unlensed spectrum is normalized
to have unit area.  Each curve is the lensed spectrum of the source at
a different lens position (not labeled), from a distance of $30R_{\rm
h}/R_\ast$ down to $0.1 R_{\rm h}/R_\ast$, the latter corresponding to the
mostly strongly magnified spectrum in each panel.  The lens position
is not evenly spaced but staggered to approximate the rate of change
in magnification.  Also note that the vertical scale is not the same
for all three panels.

As expected, significant dust content leads to a redistribution of
the intrinsic stellar flux to the infrared (top panel).  The density
of spectral curves denotes the relative change in brightness as
the event progresses.  In other words one could take cuts at various
wavelengths to obtain monochromatic lightcurves, and each one would tend
to have a different maximum amplitude (relative to the unlensed case)
and time width.  For example in the top panel, the spectra shown for
the IR ``bump'' near 10 microns are somewhat closely spaced.  Here
$R_\lambda$ is relatively large compared to $R_\ast$.  A lightcurve at
this wavelength will be broad in time and have a relatively depressed
peak amplitude because $R_\lambda \gg \hat{r}_E$ (e.g., Agol 2003).
In contrst, the spacing is more separated at NIR wavelengths.  Here
the source is more compact, and so variations between spectra vary
more rapidly as the lens transits the source.  In a light curve,
the relative amplitude would be larger and the width more
narrow in time.

\begin{figure}
\includegraphics[width=.6\textwidth]{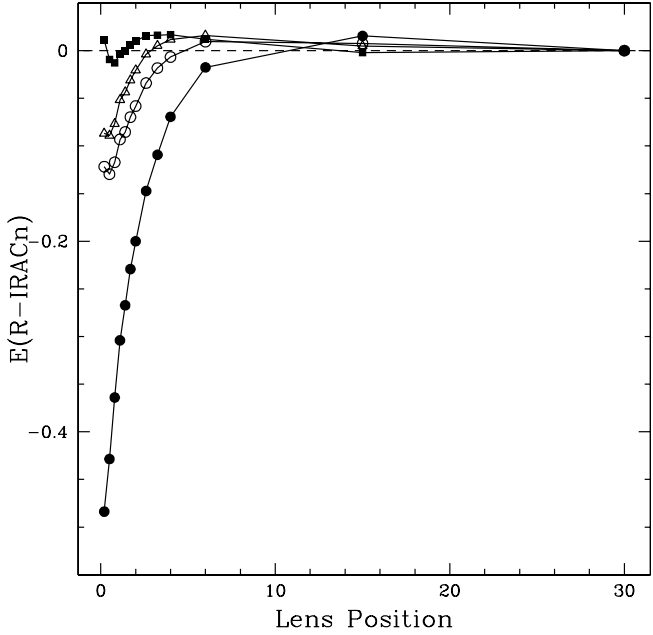}
\caption{Simulated color excesses with lens position for ground-based R-band
filter and Spitzer IRAC filters (squares for 3.6$\mu$m, triangles for
4.5$\mu$m, open circles for 5.8$\mu$m, and filled circles for 8.0$\mu$m).
The envelope optical depth is $\tau_V=1$; the condensation radius is
$R_{\rm h}=3R_\ast$; and the Einstein radius in the source plane is
$\hat{r}_E=10R_{\rm h}$.}
\label{fig4}
\end{figure}

Another way of displaying the results, and one more convenient for
observers, is a color excess lightcurve.  An example is displayed
in Figure~\ref{fig4}.  Here model spectra were integrated for an
R-band filter and for the Spitzer/IRAC filters at different lens
positions $x_{\rm lens}/R_{\rm h}$.  The excess $E(R-$IRACn), for
$n=1-4$, is relative to the unlensed value.  The R-band samples the
lightcurve for a relatively compact source at around 0.6 microns; the
IRAC $n=1$ to 4 bandpasses sample different source sizes at 3.6, 4.5, 5.8, and
8.0~$\mu$m, respectively.  As a result, the colors vary with differing
amplitudes and widths as a function of lens location relative to the star
(which of course can always be transformed to a lightcurve in time).
$E(R-$IRACn) becomes positive (``redder'') at early times because the
source is relatively larger at longer wavelengths; then peaking occurs
in the R-band as the lens nears the star because the source at that
wavelength is spatially more compact.  This information can be used to
reconstruct the extent of the source size as a function of wavelength
to test models of the density with radius in dusty winds.  These effects
will be stronger for winds of larger $\tau_V$ and
should be measurable with Spitzer if there is a ``warm'' cycle.

The question arises, what is a suitable source?  The result shown in
Figure~\ref{fig4} is for the case $\tau_V=1$, a substantial dust optical
depth at an optical wavelength.  One can relate a model optical depth to
a stellar parameters.  The optical depth is an integration through the
wind along a radius.  For $v_\infty$ the wind terminal speed of the wind,
$\sigma = \rho_{\rm dust}/\rho_{\rm wind}$ the dust to gas ratio by mass,
and $\kappa_V$ the dust opacity at 5500~\AA, the optical depth is

\begin{equation}
\tau_V = 0.001\,\sigma\,\kappa_V\,\left(\frac{\dot{M}}{10^{-9}\,
	M_\odot/{\rm yr}}\right)
	\,\left(\frac{v_\infty}{30\,{\rm km/s}}\right)^{-1}\,\left(
	\frac{R_{\rm h}}{24R_*}\right)^{-1}\,g(w),
\end{equation}

\noindent where $g(w)$ allows for a wind velocity law (e.g., if
$v/v_\infty = \sqrt{1-R_{\rm h}/r}$, then $g=2$).

For the ISM, $\sigma \approx 1\%$ and $\kappa_V \approx
200$ cm$^2$/g.  All other factors being nominal and of order a few, this
means that the wind mass-loss rate must be around $10^{-7} M_\odot/$yr for
$\tau_V$ to much exceed unity.  That level of mass loss is not achieved
in red giants but only in AGB stars.  Although rare, a couple of AGB
stars have been observed as sources in microlensing events (Gaudi,
priv.\ comm.);
consequently, there is the possibility that chromatic effects arising
from dust opacity in evolved cool star winds could be observable.
Our early results only considered the point lens case.  We can expect
even more significant chromatic variations from caustic crossing
events with binary lenses.

\section{Acknowledgements}

This project was partially funded by a partnership between the
National Science Foundation (NSF AST-0552798), Research Experiences
for Undergraduates (REU), and the Department of Defense (DoD) ASSURE
(Awards to Stimulate and Support Undergraduate Research Experiences).

\end{document}